\documentclass[letterpaper, 10 pt, conference]{ieeeconf}  

\IEEEoverridecommandlockouts                              

\usepackage{graphics} 
\usepackage{graphicx}
\usepackage{booktabs}
\usepackage{multirow}

\begin{document}

\title{\LARGE \bf
Developing and evaluating an human-automation shared control takeover strategy based on Human-in-the-loop driving simulation*

{
}

\author{Ziyao~Zhou,~Chen~Chai,~\IEEEmembership{Member,~IEEE,}~Weiru~Yin,~Xiupeng~Shi}%


\thanks{This study was jointly sponsored by the National Key Research and Development Program of China (project number: 2018YFB1600502), and the Chinese National Science Foundation (61803283).}%

\thanks{Ziyao Zhou, Chen Chai and Weiru Yin are with the College of Transportation Engineering, Tongji University and the Key Laboratory of Road and Trafﬁc Engineering of the Ministry of Education, Tongji University, Shanghai, 201804, China. (e-mail: 1833253@tongji.edu.cn; chaichen@tongji.edu.cn; 2031322@tongji.edu.cn).}%


\thanks{Xiupeng Shi is with Institute for Infocomm Research, Agency for Science Technology and Research (A*STAR), 138632, Singapore (e-mail: shix@i2r.a-star.edu.sg).}%

\thanks{Corresponding authors: Chen Chai.}
}

\maketitle
\thispagestyle{empty}
\pagestyle{empty}

\begin{abstract}

The purpose of this paper is to develop a shared control takeover strategy for smooth and safety control transition from an automation driving system to the human driver and to approve its positive impacts on drivers' behavior and attitudes. A "human-in-the-loop" driving simulator experiment was conducted to evaluate the impact of the proposed shared control takeover strategy under different disengagement conditions. Results of thirty-two drivers showed shared control takeover strategy could improve safety performance at the aggregated level, especially at non-driving related disengagements. For more urgent disengagements caused by another vehicle's sudden brake, a shared control strategy enlarges individual differences. The primary reason is that some drivers had higher self-reported mental workloads in response to the shared control takeover strategy. Therefore, shared control between driver and automation can involve driver's training to avoid mental overload when developing takeover strategies.

\end{abstract}

\section{INTRODUCTION}

Human is still required in the control loop of the automated driving system. According to the European Road Transport Research Advisory Council, highly and fully autonomous vehicles (SAE level 4 and 5) are not likely to occur by 2030, and conditionally autonomous vehicles (SAE level 3) were occurred since 2018 and will be developed entirely by 2024 \cite{dokic2015european}. Driver-vehicle interaction plays a very significant role in level 3 vehicles as the driving task is both assigned to driver and vehicle. In this context, human drivers are retained in the control loop to monitor the system and prepare for takeover requests initiated by the system, such as Level 2 or Level 3 automation system \cite{sae2014taxonomy}. They can also help the system deal with ethical troubles and legislation requirements\cite{bonnefon2016,nhtsa}. When an automated driving system failure occurs according to various causes including sensor failures or emergency traffic scenarios cannot be handled, drivers will be required to take over the control of the vehicle \cite{korber2018}. As a result, how to ensure a smooth and safe control transition from the system to the human driver at different types of system failures is the key to safety assurance of level 3 vehicles.

Takeover requests raised lots of attention in recent years, including effects on drivers’ situation awareness \cite{vlakveld2018}, drivers’ responsive times \cite{gold} and effects of warnings \cite{lylykangas2016}. Existing studies have been conducted to evaluate and improve the transition of control from the automation system to the human driver \cite{zeeb2016}. A fair amount of studies is conducted about improving the human-machine interface to enhance attention allocation before control transition and reduce reaction time \cite{van2017priming}. However, the precise evaluation of a good transition should be higher safety performance throughout the process, instead of shorter reaction time \cite{favaro2017}. Furthermore, acceptance and trust of the transition mechanism should also be a crucial issue. An untrusted and not user-friendly system may lead to improper control transitions, e.g. early hand over before requested \cite{umeno2018}.

Instead of a complete transition of control, shared control strategies combine inputs of both the human driver and the automation driving system \cite{bhardwaj2020}. One type of the shared control strategy directly blends the human input and autonomy command through an arbitrator. Drivers can feel the involvement of autonomy through observing the vehicle state after the mixed command takes effect \cite{anderson2011design}. Most studies applied the shared control strategy to improve human-vehicle interaction of Advanced Driving Assistant Systems (ADAS) \cite{nguyen2017}. Various shared control strategies, such as fictive nonlinear models, fuzzy logic models, and haptic adaptive models based on Model Predictive Control (MPC) controllers were developed \cite{nguyen2018}. These strategies are proven to improve the driving performances in terms of lane-keeping and risk avoidance through both driving simulator studies and field tests. Apart from control models to optimize shared control's driving performance, an expert driving model or reference driving model is also applied to develop the shared control strategy \cite{sekimoto}. Such methods, often called expert shared control, develop expert driving models from naturalistic driving data to obtain both control efficiency and high safety performance. Compared with other control strategies, the expert shared control strategy can have better performance for high-risk vehicle interactions.

Although most studies are conducted at the population level without considering individual differences, the shared control strategy is found to be able to improve driving performance with less visual workload and response time to driving from secondary tasks \cite{griffiths2005}. Moreover, previous studies have proved that the shared control model imposes positive effects on many aspects, including driving safety improvement and user comfort \cite{sentouh2019}.

Therefore, the objectives of this paper are: 1) develop a takeover strategy incorporating shared control, 2) evaluate the proposed shared control takeover strategy at driver’s individual level. The driving data of an actual expert driver who can accomplish the driving task with smooth vehicle motion and high safety performance was obtained. The expert shared control strategy was then developed and embedded in a "human-in-the-loop" driving simulator by combining the expert driver model with the human driver's control. A simulator experiment was then conducted on 32 participants to evaluate safety performance during the takeover process. Individual driver's mental workload, reaction time and safety performance are then evaluated.

\section{EXPERT SHARED CONTROL TAKEOVER}

\begin{figure}[t]
\flushleft
\includegraphics[width=3.2in]{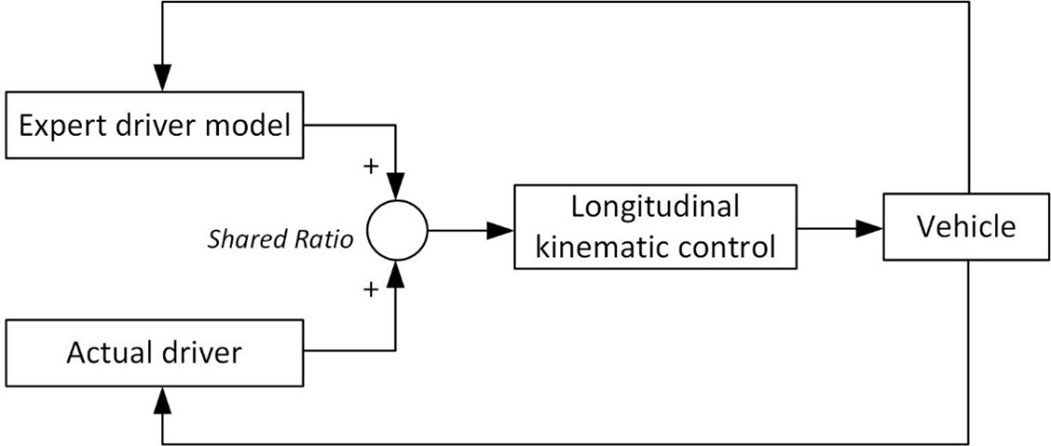}
\caption{Share control strategy.}
\label{fig_sharedcontrol}
\end{figure}

This study obtains a more realistic driving experience of Level 3 vehicles and an expert shared control strategy through replacing the autonomous driving model into an expert driving model. An experienced driver with more than ten years of driving experience was required to drive in compliance with traffic regulations, follow the lead vehicle at a steady and safe distance. The experienced driver was needed to drive several times, and the drive that well blended the safety of autonomous driving and the comfort of manual driving was selected. Gas pedal force, brake pedal force, and steering wheel angle were recorded at 20Hz. The expert driver model determines the steering wheel angle and pedal force to track the desired path recorded by the experienced driver. Euclidean distance between the real-time and recorded data in time and space get the minimal value, the behavior data at that recorded time point was chosen as the expert input at that moment, as shown in Fig.~\ref{fig_sharedcontrol}.

Through incorporating shared control in a semi-autonomous vehicle, a shared control takeover strategy can be developed. When the subject vehicle is in automated driving, the vehicle runs totally depending on the expert model. When the vehicle requires the driver to take over the control and the automation is off, the control authority is completely shifted to the driver. When the automation did not completely quit, the vehicle is controlled by the driver and automation with the respective weights (shared ratio). The shared ratio was set as 0.5, indicating the control authority is fixedly divided by driver and automation to avoid the interactive effect. The blended or independent inputs are then transferred to the simulation software through a specific Application Programming Interface (API).

\section{SIMULATOR EXPERIMENT}

\subsection{Participants}

A total of 32 participants were recruited, 22 males and 10 females. Their ages ranged from 22 to 34 years old with an average age of 25.4. The driving experience of the participants ranges from 1-11 years, with an average experience of 3.4 years. Each participant held a valid driving license. After the experiment, a cash reimbursement of 150CNY (USD\$22.00) was offered to each participant.

\subsection{Apparatus and Systems}

\begin{figure}[t]
\flushleft
\includegraphics[width=3.2in]{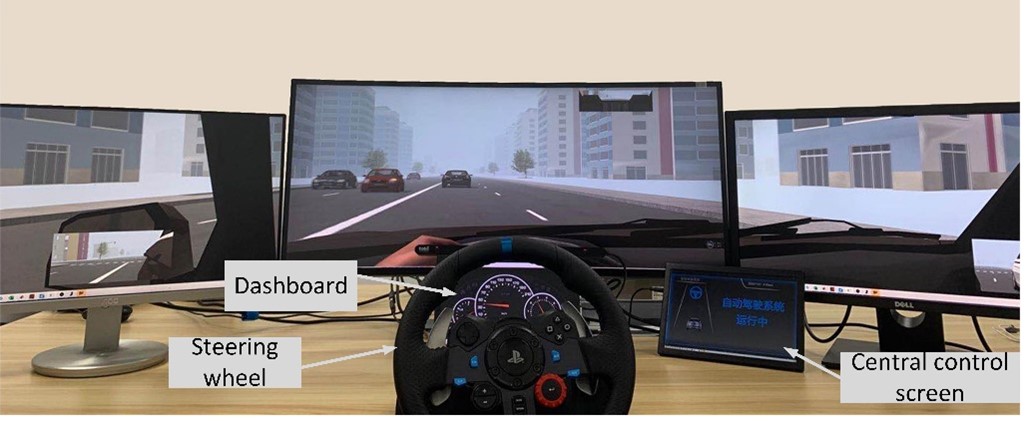}
\caption{Driving simulator.}
\label{fig_simulator}
\end{figure}

The road driving scenarios were developed using SCANeR™ studio software and projected onto three screens (23.8-inch, 34-inch curved, 23.8-inch from left to right) located 0.5m in front of the participant. The field of view was approximately 270 degrees (as shown in Fig.~\ref{fig_simulator}). The steering wheel, accelerator, and brake pedals used in this experiment were Logitech G29 Racing Wheel and Pedals. The 34-inch curved display (R=150cm) in the middle presented the central front view. In addition, two additional displays (27-inch) were used as the instrument board and central dashboard display.

\begin{figure}[t]
\flushleft
\includegraphics[width=3.2in]{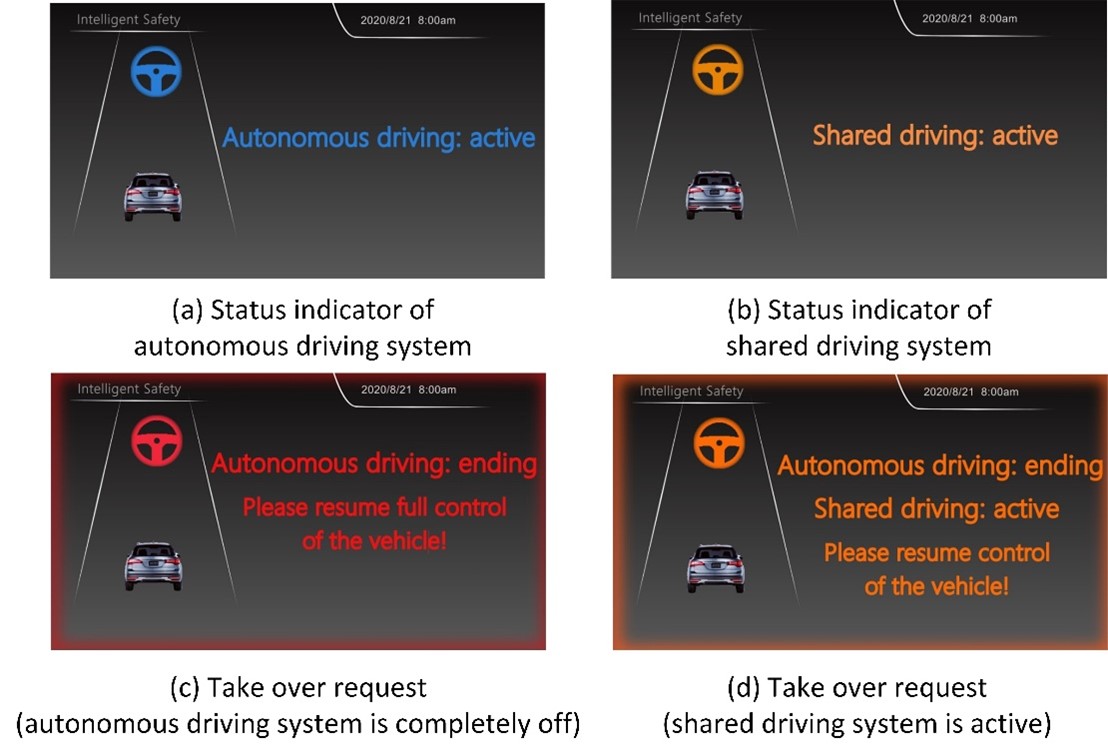}
\caption{Autonomous driving system and takeover request.}
\label{fig_tor}
\end{figure}

A 10-inch central dashboard display was employed to indicate different system status: Automation On, Shared Driving On, Automation Off, and Manual Driving On, Automation Off, and Shared Driving On. The activation of different statuses was indicated by different colors and accompanied by beeps or corresponding verbal guidance. As shown in Fig.~\ref{fig_tor}(a) and~\ref{fig_tor}(b), blue indicates the activation of autonomous driving while amber indicates shared driving. When the system was switched into the two status, a pleasant tune occurred with the texts to hint at drivers.

As shown in Fig.~\ref{fig_tor}(c) and~\ref{fig_tor}(d), the TOR interface was accompanied by beeps and verbal guidance. ‘Autonomous driving ends. Please resume full control of the vehicle’ indicated the complete exit of autonomous driving status and alerted the driver to take full control as soon as possible. ‘Shared driving is activated. Please resume control of the vehicle’s alerted the driver to take control and cooperative with the vehicle in time for safe driving.

\subsection{Experimental Design}

\begin{figure}[t]
\flushleft
\includegraphics[width=2.8in]{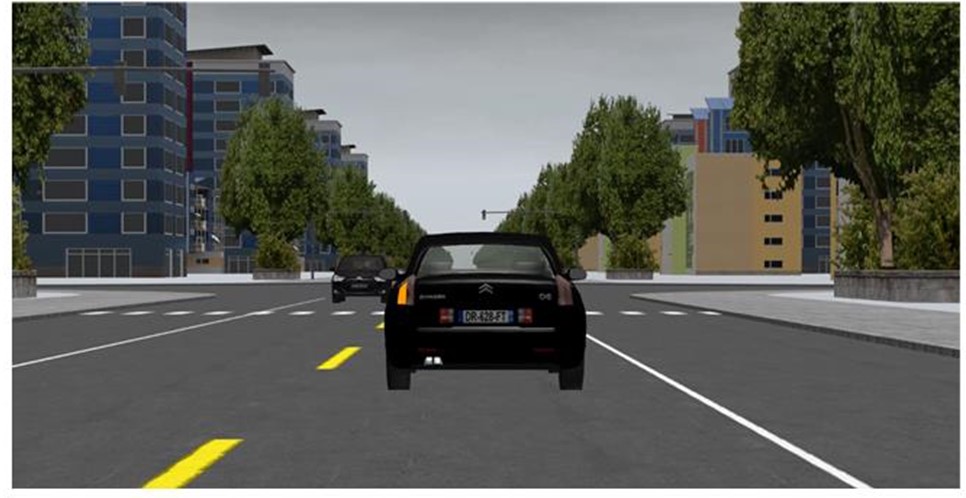}
\caption{Driving environment.}
\label{fig_environment}
\end{figure}

The designed driving environment was an urban road with two lanes of traffic in each direction in the misty daytime (Fig.~\ref{fig_environment}). Surrounding vehicles were programmed to drive at 30-50 km/h. The lead vehicle employed an automated driving algorithm, which enabled sign and emergency observing.

Participants were instructed to follow a lead vehicle at a safe distance. At the beginning of each drive, autonomous driving was performed, and the driver should monitor the surrounding environment for safety. During the driving process, the autonomous driving system would fail and exit at specific positions. Then a safety-critical event occurred not far ahead. Drivers needed to take control in time to avoid a potential forward collision.

\begin{figure}[t]
\flushleft
\includegraphics[width=2.8in]{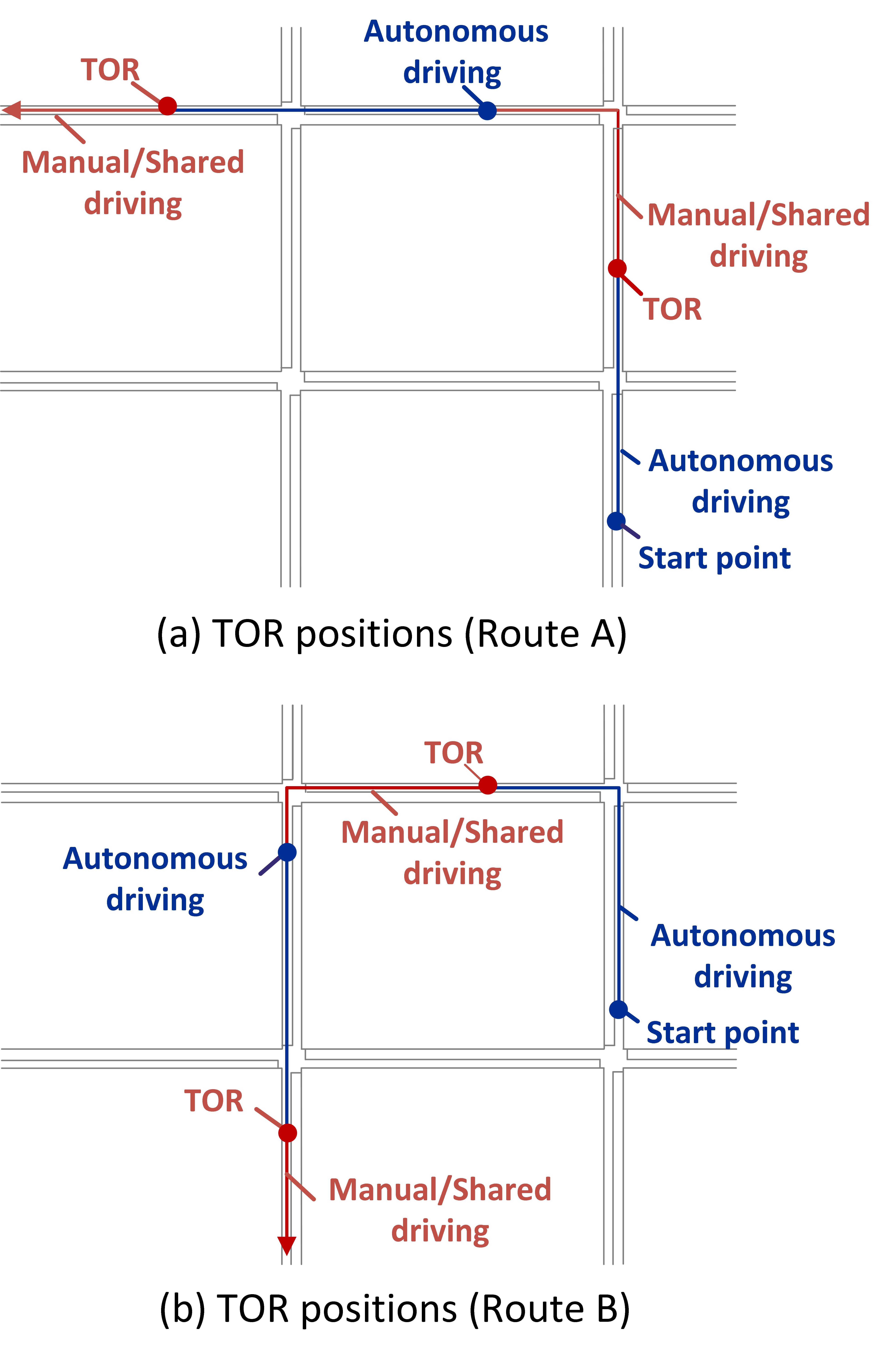}
\caption{Designed driving routes.}
\label{fig_route}
\end{figure}

Each participant had five drives. As is shown in Fig.~\ref{fig_route}, the driving routes were shown in lines with an arrow. At the beginning of each drive, the automation was on. TOR occurred at the red points, and the system state was switched to manual/shared driving. When drivers took over the vehicle to avoided the collision, the system would reenter the autonomous driving again after a steadily car-following status is observed. The positions of blue points were not fixed but based on the judgment of the vehicle operating status. Each drive included two TOR events. The differences between the two driving routes in Fig.~\ref{fig_route}(a) and Fig.~\ref{fig_route}(b) were the TOR events' positions. Driving routes in five drives were chosen at a random order to prevent drivers from forecasting emergency.

Two kinds of disengagement events were designed in the experiment. The first type is a sudden disengagement due to non-traffic related reasons (e.g. sensor failure). The failure triggered TOR immediately. To assess driver’s collision avoidance ability after takeover, a sudden brake of the lead vehicle at –9.8 m/$s^2$ would be occurred 1s after TOR. The second TOR was due to traffic-related reasons, as TOR and the sudden brake happened simultaneously because the situation exceeded the system capacity. In this study, the first type of disengagement is called “ordinary scenario”, and the second type is called “urgent scenario”.

\begin{figure}[t]
\flushleft
\includegraphics[width=3.2in]{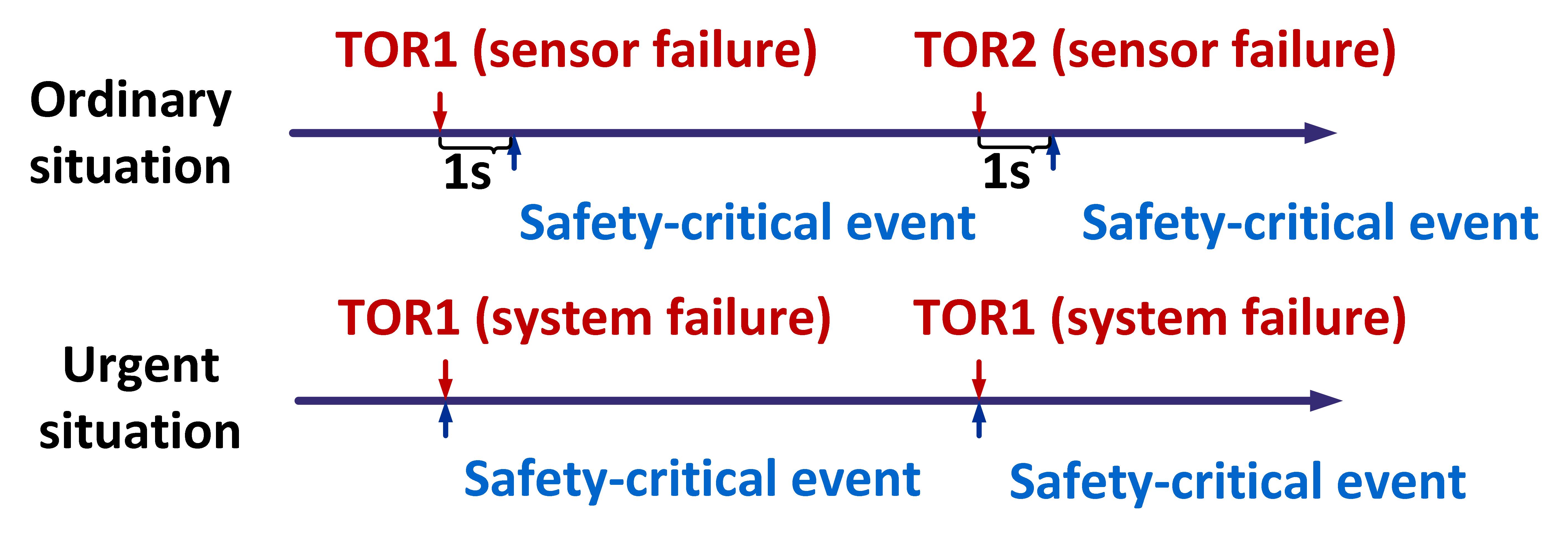}
\caption{Descriptions of takeover requests.}
\label{fig_description}
\end{figure}

\begin{figure}[t]
\flushleft
\includegraphics[width=3.2in]{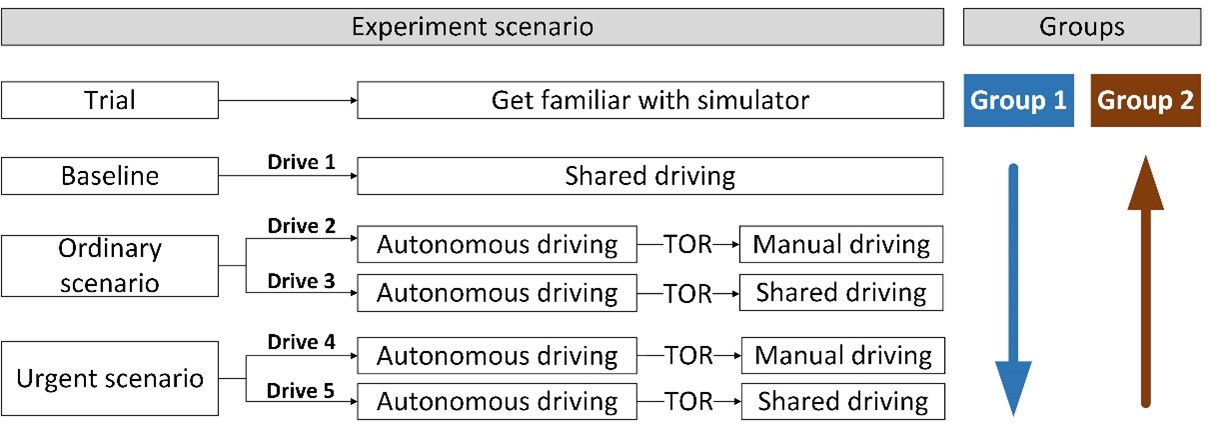}
\caption{Experiment process.}
\label{fig_process}
\end{figure}

Thirty-two participants were divided into two groups, equally and randomly. T-tests of age, gender, and driving experience have been conducted. There were no significant differences between the two groups. The participants were informed to drive as normal in daily life and comply with traffic rules as in real-driving situations. Before the experiment, each driver had about a 5-minute trial session to get familiar with the driving simulator operations and the autonomous driving state. During each drive of the formal experiment, each participant drove five times with two TORs. Different scenarios were presented in a counterbalanced order for two groups to reduce experiment order effects such as fatigue or learning (Fig.~\ref{fig_process}). Each drive lasted about 4 min. After each drive, participants filled in a NASA Task Load Index (TLX) questionnaire to rate their subjective workload during the driving process (Hart and Staveland, 1988).

\subsection{Measurements}

\begin{figure}[t]
\flushleft
\includegraphics[width=3.2in]{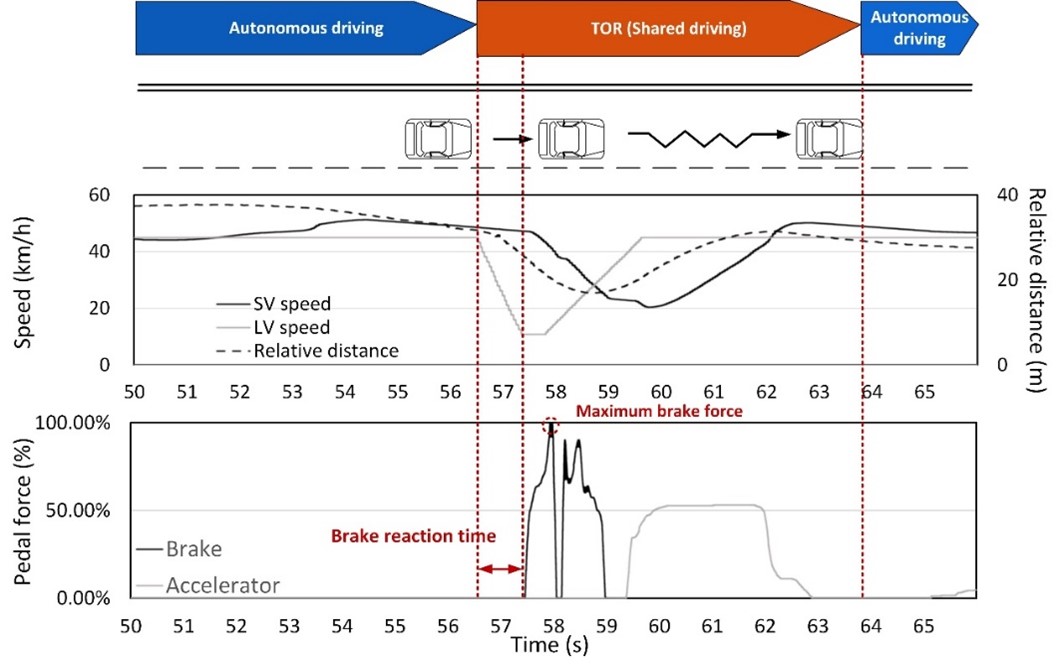}
\caption{Computation of brake reaction time.}
\label{fig_breakreaction}
\end{figure}

In this study, the experiment was designed as a 2 (urgent/ordinary scenarios)×2 (takeover strategies) within-subjects repeated measures. Dependent variables included both objective (i.e., reaction time) and subjective measures (i.e., subjective ratings of workload). Detailed definitions were as follows:

1) Reaction time (RT): time is taken to press the brake pedal after TOR was prompted, as shown in Fig.~\ref{fig_breakreaction}.

\begin{figure}[t]
\flushleft
\includegraphics[width=3.2in]{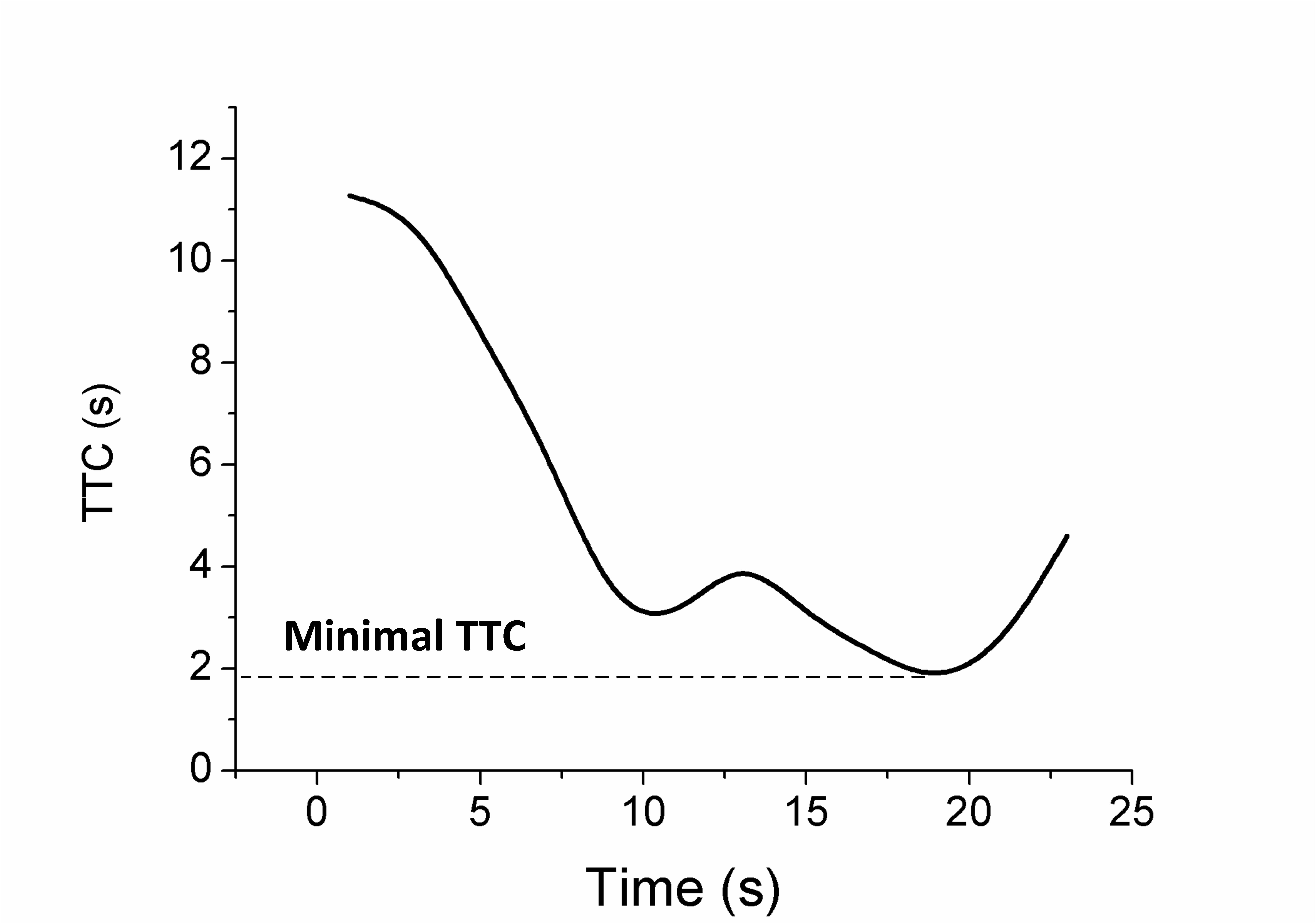}
\caption{Computation of minimal TTC.}
\label{fig_ttc}
\end{figure}

2) Safety indicators: Three safety indicators are computed according to the relative vehicle movements of the subject and lead vehicles. Minimal time to Collision (minTTC) is computed as the minimal value of time to a potential collision between two vehicles after TOR, as shown in Fig.~\ref{fig_ttc}. Time Exposed TTC (TET) were derived from TTC to evaluate the dynamic potential risk. It is computed as the duration of TTC smaller than a certain TTC threshold, as shown in Fig.~\ref{fig_ttc}. Time Exposed TTC (TET) were derived from TTC to evaluate the dynamic potential risk. It is computed as the duration of TTC smaller than a certain TTC threshold, as shown in Figure 8b. Similarly, Time Integrated TTC (TIT) was derived from TTC to evaluate the integration of TTC smaller than a certain threshold.

3) Subjective ratings of workload: NASA-TLX requires participants to rate the demands in mind, physics, time, frustration, effort, and performance. The overall workload index is computed by averaging the scores of six dimensions in this study.

\section{RESULTS AND DISCUSSIONS}

\subsection{Safety performance of shared control takeovers}

\begin{figure}[t]
\flushleft
\includegraphics[width=3.2in]{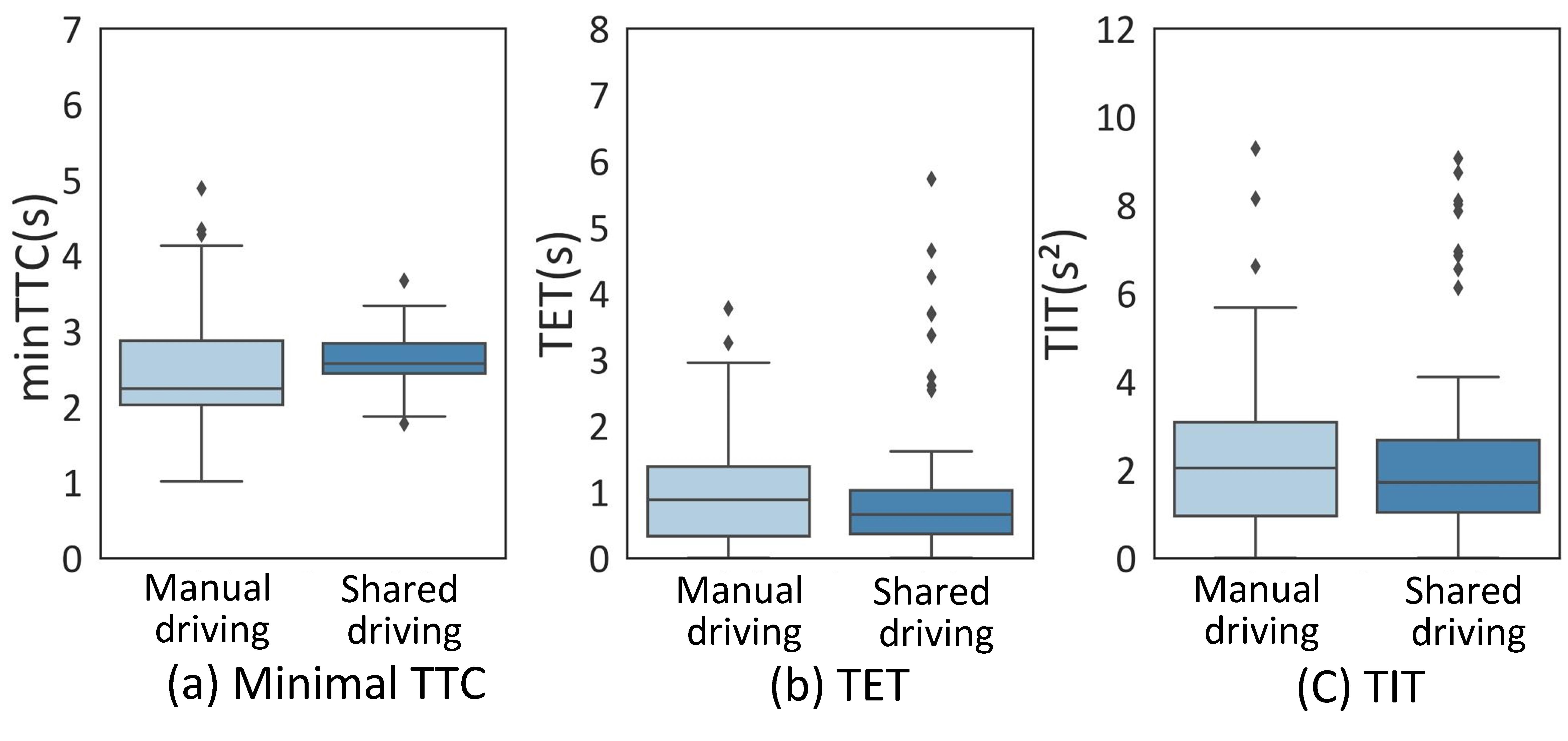}
\caption{Safety performance of shared and manual takeover in ordinary scenario.}
\label{fig_ordinary}
\end{figure}

\begin{figure}[t]
\flushleft
\includegraphics[width=3.2in]{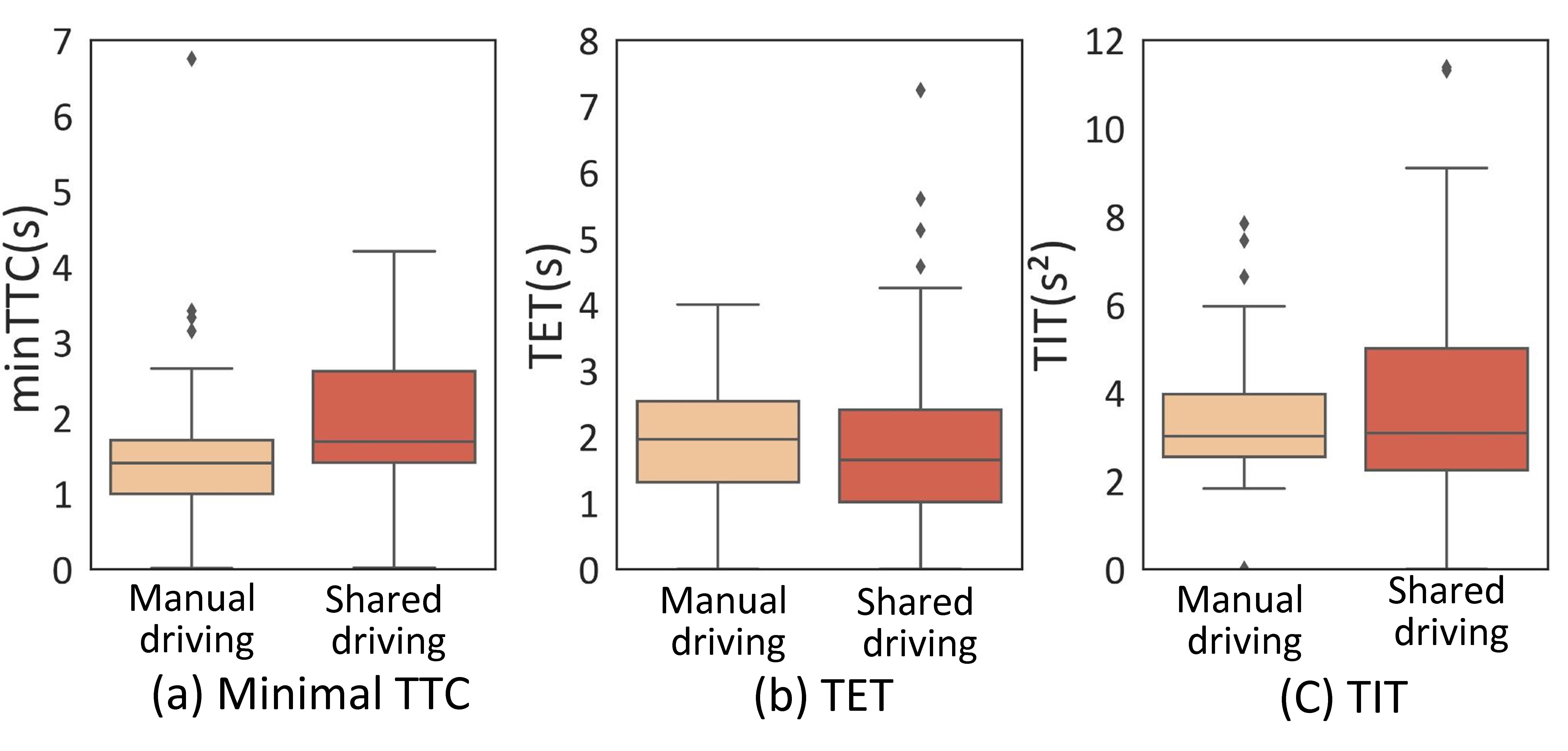}
\caption{Safety performance of shared and manual takeover in urgent scenario.}
\label{fig_urgent}
\end{figure}

Fig.~\ref{fig_ordinary} and Fig.~\ref{fig_urgent} summarized distributions of safety indicators under ordinary and urgent scenarios, respectively. According to Figures 10a and 11a, the minTTC of shared control takeover was larger than manual driving under both kinds of scenarios. The decrease was significant under 5\% significance level (p=0.024 under ordinary scenarios and p=0.044 under urgent scenarios). The results indicated that the proposed shared control take over strategy effectively improves safety performance by preventing severe conflict.
Significant TET and TIT reduction between shared control and manual takeover were not observed at 5\% level. However, the share control takeover's effectiveness under urgent scenarios is found to be largely affected by individual differences. TET and TIT's interquartile ranges were larger in shared driving than that in manual driving compared with ordinary scenarios. To further evaluate the individual differences, individual drivers' vehicle movements and safety performances were extracted and analyzed.

\subsection{Individual differences on takeover reactions}

\begin{figure}[t]
\flushleft
\includegraphics[width=3.2in]{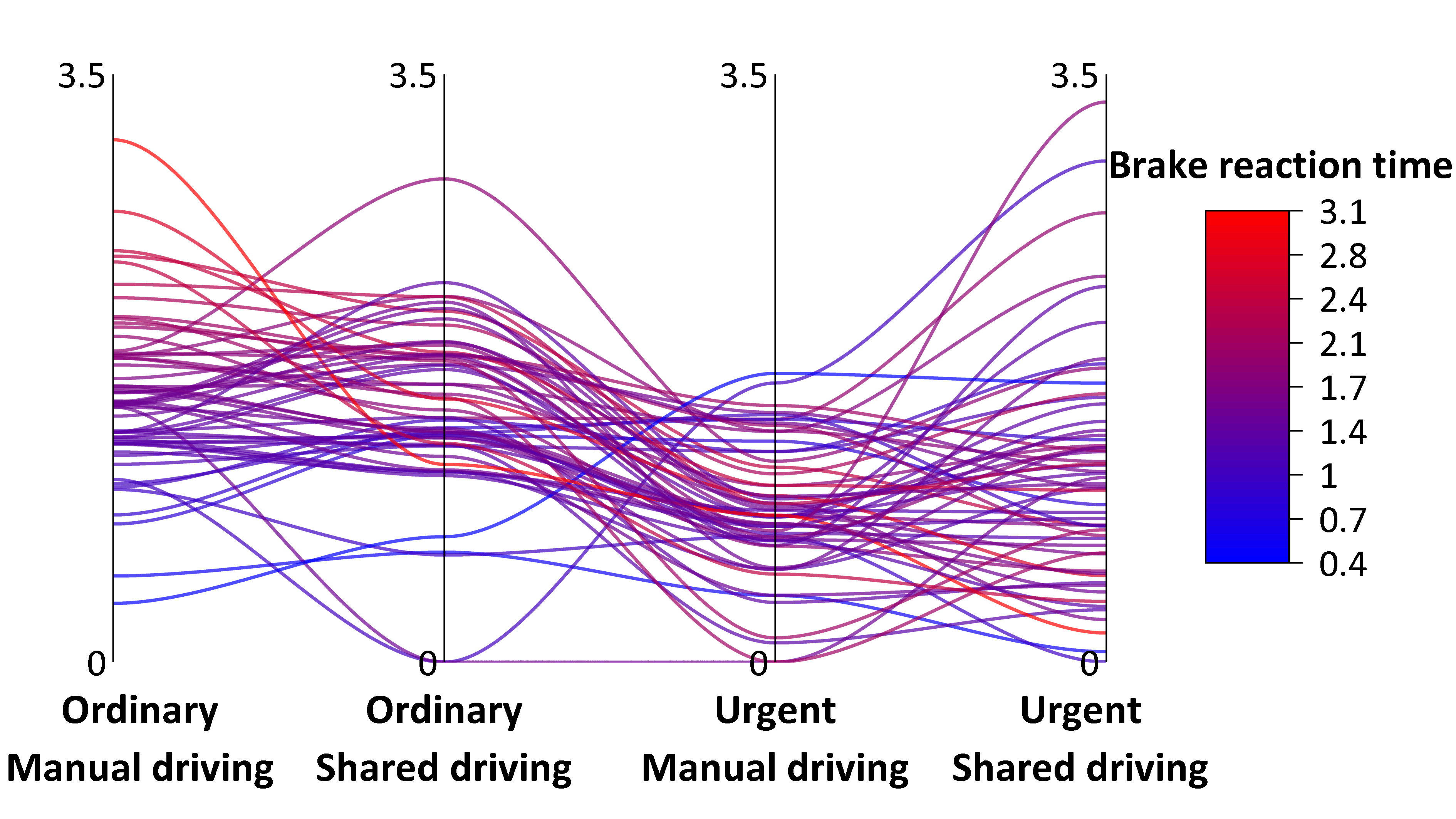}
\caption{Individual brake reaction time under tested scenarios.}
\label{fig_individual}
\end{figure}

Fig.~\ref{fig_individual} shows the brake reaction time of each participant mainly affected by individual differences. Four vertical axles represent reaction times under four driving scenarios. In general, brake reaction time is longer in shared control takeover compared to manual takeover. This can be explained as drivers may delay their braking response with a shared driving system's assistance.

\begin{figure}[t]
\flushleft
\includegraphics[width=3.2in]{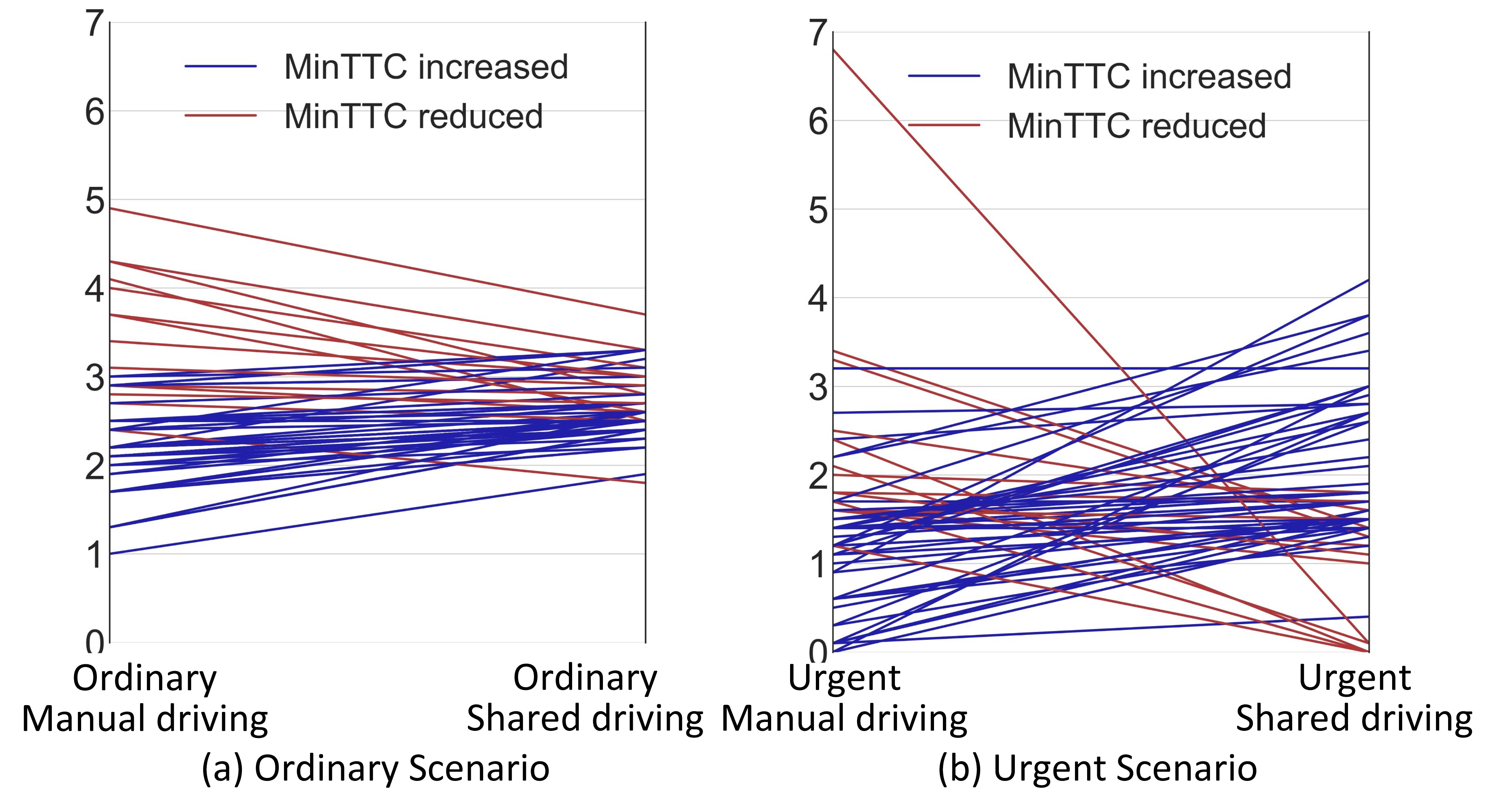}
\caption{Drivers classified by different impacts on minTTC.}
\label{fig_classifybyttc}
\end{figure}

Therefore, further evaluation of individual differences is analyzed by comparing each driver's behavior with and without shared control takeover. Firstly, Fig.~\ref{fig_classifybyttc} shows drivers classified by minTTC differences between shared control takeover and manual takeover. Red lines represent drivers with smaller minTTC (worse safety performance) in shared control takeover, while blue lines represent drivers with larger minTTC (better safety performance) in shared control takeover. 

From Fig.13a, it can be found that the shared control takeover can reduce individual differences under ordinary scenarios. A relatively safe and appropriate minimal TTC can be achieved when driver needed to take over half control of the vehicle. However, the shared control takeover didn't work for everyone when it comes to urgent situations. It is noteworthy that several drivers were involved in crashes when the shared control takeover was introduced, as shown in Fig.13b. 

To further analyze drivers’ individual differences under urgent takeover scenarios. Drivers are divided into several groups by comparing the differences between manual takeover and shared control takeover, as summarized in Tab.~\ref{tab:table1}. For each driver, minTTC difference was calculated by subtracting minimal TTC under manual takeover from that under shared control takeover. The positive values indicate that the safety performance under shared control takeover is better than that under manual takeover, and vice versa. Drivers were then divided into three groups taking -0.2 and 0.2 seconds of minTTC difference as critical thresholds\ cite. The first group is performed better with manual takeover, the 2nd group performed similar under two takeover strategies, and the 3rd group performed better after shared control strategy. T-tests were conducted significant differences are marked as asterisks (*, between 1$^st$ and 2$^nd$ groups) and pounds (\#, between 2$^nd$ and 3$^rd$ groups) at 5\% significant level.

As shown in Tab.1, under urgent takeover scenarios, significantly differences are observed from reaction times and self-reported workloads. The driver with worse performance after shared control takeover reports significantly higher mental workload. Moreover, although the driver reacted quicker than in manual takeover, the deceleration was not efficient and thus result in a worse safety performance. On the other hand, drivers with higher performance after shared control takeover reports significantly lower mental workload and smaller reaction time. Last but not least, all the females were in this group, indicating the shared control takeover was more suitable for female drivers.

\begin{table}[]
\caption{Group differences under urgent takeover scenarios}
\label{tab:table1}
\begin{tabular}{@{}lcccl@{}}
\toprule
\multirow{2}{*}{}            & \multicolumn{3}{c}{Minimal TTC difference}             &  \\ \cmidrule(l){2-5} 
                             & \textless{}-0.2s & -0.2$\sim$0.2s & \textgreater{}0.2s &  \\ \cmidrule(r){1-1}
No.of   drivers              & 1                & 5              & 26                 &  \\
Reaction   time difference*\# & -0.8             & 0.25 (0.86)    & 0.07 (0.50)        &  \\
Workload   difference*\#      & 18.67            & 6.13 (9.61)    & 0.19 (13.18)       &  \\
Age                          & 22               & 27.50 (4.36)   & 25.42 (2.77)       &  \\
Driving   Experience         & 3                & 6.25 (3.40)    & 4.23 (2.39)        &  \\
Gender\#                      & M             & 5M        & 16M 10F  &  \\ \bottomrule
\end{tabular}
\end{table}

 illustrates the existence of individual differences. The reaction time under urgent situations were scattered, ranging from 0 to 3.3 seconds. It is hard to observe a common trend that can summarize the effect of shared control takeover.

\begin{figure}[t]
\flushleft
\includegraphics[width=3.2in]{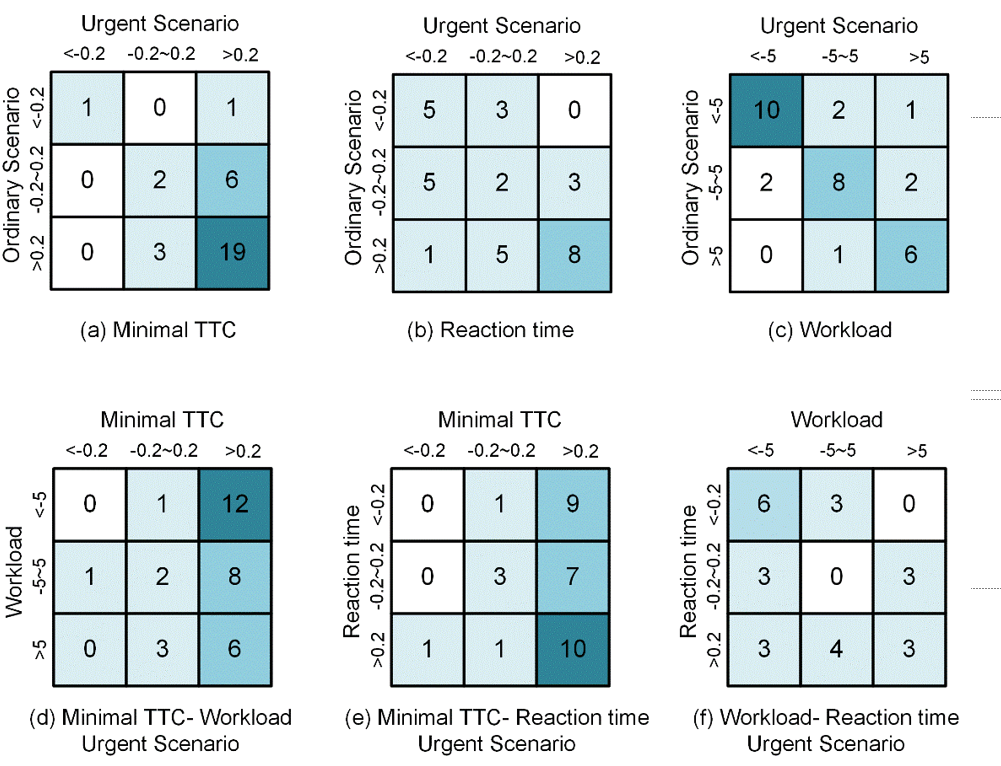}
\caption{Drivers classified by different impacts on minTTC.}
\label{fig_square}
\end{figure}

Furthermore, the consistency between ordinary and urgent scenarios in the aspects of minimal TTC difference, reaction time and workload. As shown in Fig.~\ref{fig_square},the squares were painted from shallow to dark blue depending on the increase of driver numbers in corresponding group. It can be found that the minimal TTC difference and workload difference showed a common tendency in the two scenarios. Most drivers can increase their minimal TTC more than 0.2 seconds under the shared control takeover. Also, driver workload was more affected by the takeover strategy instead of the scenario urgency. Figures 14d-f demonstrated the relationship between minimal TTC difference, reaction time and workload under urgent scenario. It is found that the decrease of driver workload was along with the increase of minimal TTC. The minimal TTC was extended among most drivers under urgent scenarios, but their reaction time was scattered. Also, there was no obvious association between workload and reaction time.

\section{CONCLUSIONS}

In this paper, an expert shared control takeover strategy of semi-autonomous vehicle is proposed and evaluated at individual level. The strategy aimed to reduce driver’s mental workload and improving the safety performance through a smooth control transition between automation driving system to the human driver. A “human-in-the-loop” driving simulator study is conducted through embedding the shared control takeover strategy based on co-simulation of SCANeR and Matlab Simulink. 
Experimental results of 32 drivers showed shared control takeover strategy can improve safety performance at group level. This is consistent with many previous studies of shared control driving and shared control takeover. 
Further evaluation on individual difference have found that the shared control strategy can achieve a unified better safety performance at non-traffic related disengagements. For more urgent disengagements, shared control strategy enlarges individual differences. The results also showed that worse safety performance is related with higher mental workloads caused by shared control strategy rather than disengagement types. This study confirmed the effectiveness of the proposed expert shared control strategy on improving safety performance. Controlling driver's mental overload through instructions and training are also necessary in further studies.

\addtolength{\textheight}{-12cm}   






\bibliographystyle{IEEEtran}
\bibliography{X}

\begin{thebibliography}{10}
\providecommand{\url}[1]{#1}
\csname url@samestyle\endcsname
\providecommand{\newblock}{\relax}
\providecommand{\bibinfo}[2]{#2}
\providecommand{\BIBentrySTDinterwordspacing}{\spaceskip=0pt\relax}
\providecommand{\BIBentryALTinterwordstretchfactor}{4}
\providecommand{\BIBentryALTinterwordspacing}{\spaceskip=\fontdimen2\font plus
\BIBentryALTinterwordstretchfactor\fontdimen3\font minus
  \fontdimen4\font\relax}
\providecommand{\BIBforeignlanguage}[2]{{%
\expandafter\ifx\csname l@#1\endcsname\relax
\typeout{** WARNING: IEEEtran.bst: No hyphenation pattern has been}%
\typeout{** loaded for the language `#1'. Using the pattern for}%
\typeout{** the default language instead.}%
\else
\language=\csname l@#1\endcsname
\fi
#2}}
\providecommand{\BIBdecl}{\relax}
\BIBdecl

\bibitem{dokic2015european}
J.~Dokic, B.~M{\"u}ller, and G.~Meyer, ``European roadmap smart systems for
  automated driving,'' \emph{European Technology Platform on Smart Systems
  Integration}, vol.~39, 2015.

\bibitem{sae2014taxonomy}
``Taxonomy and definitions for terms related to on-road motor vehicle automated
  driving systems.sae on-road automated vehicle standards committee and
  others,'' \emph{SAE Standard J}, vol. 3016, pp. 1--16, 2014.

\bibitem{bonnefon2016}
J.-F. Bonnefon, A.~Shariff, and I.~Rahwan, ``The social dilemma of autonomous
  vehicles,'' \emph{Science}, vol. 352, 06 2016.

\bibitem{nhtsa}
F.~A.~V. Policy, ``Accelerating the next revolution in roadway safety, national
  highway traffic safety administration, department of transportation,
  washington, dc.'' \emph{Transportation}, 2016.

\bibitem{korber2018}
M.~Körber, L.~Prasch, and K.~Bengler, ``Why do i have to drive now? post hoc
  explanations of takeover requests,'' \emph{Human Factors The Journal of the
  Human Factors and Ergonomics Society}, vol.~60, p. 305–323, 05 2018.

\bibitem{vlakveld2018}
W.~Vlakveld, N.~Nes, J.~de~Bruin, L.~Vissers, and M.~Kroft, ``Situation
  awareness increases when drivers have more time to take over the wheel in a
  level 3 automated car: A simulator study,'' \emph{Transportation Research
  Part F: Traffic Psychology and Behaviour}, vol.~58, pp. 917--929, 10 2018.

\bibitem{gold}
C.~Gold, D.~Dambock, L.~Lorenz, and K.~Bengler, ``"take over!" how long does it
  take to get the driver back into the loop?'' \emph{Proceedings of the Human
  Factors and Ergonomics Society Annual Meeting}, vol.~57, pp. 1938--1942, 09
  2013.

\bibitem{lylykangas2016}
J.~Lylykangas, V.~Surakka, K.~Salminen, A.~Farooq, and R.~Raisamo, ``Responses
  to visual, tactile and visual–tactile forward collision warnings while gaze
  on and off the road,'' \emph{Transportation Research Part F: Traffic
  Psychology and Behaviour}, vol.~40, pp. 68--77, 07 2016.

\bibitem{zeeb2016}
K.~Zeeb, A.~Buchner, and M.~Schrauf, ``Is take-over time all that matters? the
  impact of visual-cognitive load on driver take-over quality after
  conditionally automated driving,'' \emph{Accident; analysis and prevention},
  vol.~92, pp. 230--239, 04 2016.

\bibitem{van2017priming}
R.~M. van~der Heiden, S.~T. Iqbal, and C.~P. Janssen, ``Priming drivers before
  handover in semi-autonomous cars,'' in \emph{Proceedings of the 2017 CHI
  conference on human factors in computing systems}, 2017, pp. 392--404.

\bibitem{favaro2017}
F.~Favarò, S.~Eurich, and N.~Nader, ``Autonomous vehicles' disengagements:
  Trends, triggers, and regulatory limitations,'' \emph{Accident; analysis and
  prevention}, vol. 110, pp. 136--148, 11 2017.

\bibitem{umeno2018}
R.~Umeno, M.~Itoh, and S.~Kitazaki, ``Influence of automated driving on
  driver’s own localization: a driving simulator study,'' \emph{Journal of
  Intelligent and Connected Vehicles}, vol.~1, 11 2018.

\bibitem{bhardwaj2020}
A.~Bhardwaj, A.~Ghasemi, Y.~Zheng, H.~Febbo, P.~Jayakumar, T.~Ersal, J.~Stein,
  and B.~Gillespie, ``Who’s the boss? arbitrating control authority between a
  human driver and automation system,'' \emph{Transportation Research Part F:
  Traffic Psychology and Behaviour}, vol.~68, pp. 144--160, 01 2020.

\bibitem{anderson2011design}
S.~J. Anderson, S.~C. Peters, T.~E. Pilutti, and K.~Iagnemma, ``Design and
  development of an optimal-control-based framework for trajectory planning,
  threat assessment, and semi-autonomous control of passenger vehicles in
  hazard avoidance scenarios,'' in \emph{Robotics Research}.\hskip 1em plus
  0.5em minus 0.4em\relax Springer, 2011, pp. 39--54.

\bibitem{nguyen2017}
A.-T. Nguyen, C.~Sentouh, and J.-C. Popieul, ``Driver-automation cooperative
  approach for shared steering control under multiple system constraints:
  Design and experiments,'' \emph{IEEE Transactions on Industrial Electronics},
  vol.~64, pp. 3819--3830, 05 2017.

\bibitem{nguyen2018}
------, ``Sensor reduction for driver-automation shared steering control via an
  adaptive authority allocation strategy,'' \emph{IEEE/ASME Transactions on
  Mechatronics}, vol.~23, pp. 5--16, 02 2018.

\bibitem{sekimoto}
S.~Yuki, I.~Shintaro, and R.~Pongsathorn, ``Adaptive haptic shared control in
  steering operation for curved path tracking assistance,'' in
  \emph{Proceedings of 5th International Symposium on Future Active Safety
  Technology Towards Zero-Traffic-Accidents (FAST-zero’19)}, 2019.

\bibitem{griffiths2005}
P.~Griffiths and B.~Gillespie, ``Sharing control between humans and automation
  using haptic interface: Primary and secondary task performance benefits,''
  \emph{Human factors}, vol.~47, pp. 574--90, 10 2005.

\bibitem{sentouh2019}
C.~Sentouh, A.-T. Nguyen, A.~Benloucif, and J.-C. Popieul, ``Driver-automation
  cooperation oriented approach for shared lateral control design,'' \emph{IEEE
  Transactions on Control Systems Technology}, vol.~27, pp. 1962--1978, 09
  2019.

\end{thebibliography}

\end{document}